\documentclass[final,twocolumn,3p,times]{elsarticle}
\usepackage{graphicx}
\usepackage{amssymb}
\usepackage{amsmath}
\usepackage{hyperref}
\usepackage{color}

\journal{Physica E (and accepted)}

\begin{document}
 \begin{frontmatter}

\title{Stable multi-peak   vector  solitons in  spin-orbit coupled 
       spin-$1$ polar condensates}

\author[ift]{Sadhan K. Adhikari\corref{author}}
\ead{sk.adhikari@unesp.br}
\cortext[author]{Corresponding author}

\address{Instituto de F\'{\i}sica Te\'orica, Universidade Estadual
             Paulista - UNESP, \\ 01.140-070 S\~ao Paulo, S\~ao Paulo, Brazil}
      

\begin{abstract}
We demonstrate the formation of  multi-peak 
three-component   stationary stripe  vector solitons in a 
quasi-one-dimensional spin-orbit-coupled hyper-fine spin $F=1$  polar 
Bose-Einstein 
condensate.  
The present  investigation is carried out through a numerical solution by imaginary-time propagation
and an analytic variational approximation 
of the underlying mean-field Gross-Pitaevskii equation. Simple  analytic results for energy and component densities were 
found to be in excellent agreement with the numerical results for solitons with more than 100 pronounced maxima and minima. The vector solitons are one of the two types: dark-bright-dark
or bright-dark-bright.  In the former a maximum density in   component $F_z=0$ at the center  is accompanied by a zero in components $F_z=\pm 1$. The opposite happens in the latter case. 
The vector  solitons are demonstrated to be mobile 
and dynamically stable.     The collision between two such vector solitons is found to be quasi elastic at large velocities 
with the conservation of total density of each vector soliton. 
However, at very small velocity, the collision is inelastic   with a destruction of the initial vector solitons.  It is possible to observe and study the predicted SO-coupled
vector solitons in a laboratory.

\end{abstract}


\end{frontmatter}

\section{Introduction}
\label{Sec-I}

A soliton or solitary wave is a self-reinforcing wave packet that maintains its shape while propagating at a constant velocity. Solitons are created by a cancellation of nonlinear and dispersive effects in the medium. 
 Solitons  have been observed \cite{Kivshar}  in  water waves, in non-linear optics, and in  
Bose-Einstein condensates (BECs)  among others. 
Solitons have been generated  in a BEC of $^7$Li \cite{li} and $^{85}$Rb \cite{rb} atoms
 by  manipulating  the  non-linear atomic attraction  near a 
Feshbach resonance \cite{Inouye}. 
Solitons have also been studied in  binary 
BEC mixtures  \cite{Perez-Garcia}. Properties of BEC solitons are well described \cite{vmpg} by  the mean-field 
Gross-Pitaevskii (GP) equation \cite{gp}.

The experimental realization of  a spinor BEC of $^{23}$Na atoms with hyper-fine spin 
$F=1$ \cite{exp}   required a 
 generalization of the GP equation \cite{Ohmi}, which  can   describe its properties.  
In a spinor BEC of neutral atoms, 
there is no  natural spin-orbit (SO) coupling.  However, 
 a synthetic   SO coupling can  realized in a spinor BEC     by a management of 
external electromagnetic fields  \cite{stringari,rev}. Different managements are possible which generate  different types of  SO coupling 
between spin and momentum  in   a spinor BEC.   Two such possible   
SO couplings are due to  Rashba \cite{Rashba} and 
Dresselhaus \cite{Dresselhaus}.  An equal-strength  mixture of Rashba and Dresselhaus  SO couplings was first realized experimentally in a  pseudo spin-1/2 spinor BEC  formed of   
two ($F_z=0$ and $-1$) of the three  hyper-fine spin components of    
the $F = 1$ state 5S$_{1/2}$ of $^{87}$Rb \cite{Lin}. The coupling scheme of this truncated two-component spinor BEC is quite similar to that of a spin-1/2 particle and can be described by the Pauli spin-1/2 spinors. This is why this dynamics is ususlly termed pseudo-spin-1/2 dynamics.  
Similar SO-coupled BECs   were   formed later and studied in different 
laboratories \cite{diff}.
Different possible SO couplings in spinor BECs and 
the ways to engineer  these in a laboratory are addressed in review articles   \cite{stringari}.  
Possible ways of realizing the SO coupling in three-component spin-1 BEC have been discussed \cite{SOspin1}.

 Solitons have been 
extensively studied in spinor BECs without SO coupling \cite{Ieda}.  
Solitonic structures in SO-coupled pseudo-spin-$1/2$ \cite{rela, stripe},  spin-$1$  
\cite{Liu,sol1d}  and spin-2 \cite{SO-sol-spin2} BECs have also been investigated theoretically. These studies were extended to quasi-solitons confined in two \cite{sol2d}
and three \cite{sol3d} dimensions. 
 Different types of SO coupling introduce rich dynamics through
different types of derivative couplings among the component wave functions of the mean-field model.

A spin-$1$ spinor BEC is controlled  by two interaction strengths, e.g.,
  $c_0\propto (a_0+2a_2)/3$ and $c_2 \propto (a_2-a_0)/3$, with $a_0$ and 
$a_2$ the scattering lengths in total spin $0$ and 
2 channels,  respectively \cite{Ohmi}. All  spin-1 spinor BECs can be classified into  two distinct types \cite{Ohmi,stringari}: 
a ferromagnetic   BEC ($c_2<0$) and an anti-ferromagnetic or a polar BEC ($c_2>0$).  
Most of the previous studies on SO-coupled BEC soliton \cite{rela}
were limited to the  pseudo-spin-1/2 case, which does and can not exhibit  the full rich dynamics of  a spin-1 spinor. Like its three-component counterpart, a pseudo-spin-1/2 spinor does not have the distinct varieties: ferromagnetic and polar. For example, an SO-coupled spin-1 
polar soliton is a stripe multi-peak one, whereas the ferromagnetic soliton is a single-peak  one \cite{sol1d}. 

 { In this paper}, we study    vector solitons in 
a SO-coupled spin-$1$ polar  BEC   using a mean-field  coupled  GP 
equation. 
We consider the SO coupling  $(\propto \gamma  p_x \Sigma_y)$, which is an equal-strength  mixture of Rashba and Dresselhaus SO couplings realized in the 
pioneering experiment \cite{Lin}.
 Here $ p_x$ is the $x$ component of momentum, $\gamma$ the strength of SO coupling, and $\Sigma_y$ is the $y$ component of the spin-1   matrix $\Sigma$.  
With this SO coupling, we study the properties of  
three-component vector solitons in a   one-dimensional (1D)  polar BEC along the $x$  axis   \cite{sol1d}.    The polar solitons generated with 
the SO coupling 
  $(\propto \gamma  p_x \Sigma_y)$ have identical density distribution  and energy as the solitons with the SO coupling 
  $(\propto \gamma  p_x \Sigma_x)$. Because of the rotational symmetry in spin space 
around the symmetry  axis $z$, these two SO couplings are equivalent.  The
generated  vector solitons can have a very large number of pronounced maxima and minima along the spatial direction and hence are stripe vector solitons \cite{stripe}.  We demonstrate the formation of stable robust stationary vector solitons, with more than 100 maxima and minima, numerically  by imaginary-time propagation.

Simple analytic expression  for the densities  and energies of the vector soliton were obtained from the  
 SO-coupled GP equation employing  a plausible   approximation and a  variational scheme which minimize the energy functional.  It is remarkable that the analytic results are independent of the  interaction strength  $c_2$ determined 
only by the strength $c_0$.   The analytic results for the density and energy of the SO-coupled polar vector soliton 
are found to be in excellent agreement with the numerical results of the same obtained by imaginary-time propagation. The numerical results, however, has a weak dependence on the interaction strength $c_2$.

 We also study the dynamics of the vector soliton numerically by real-time simulation. The dynamical stability of the vector soliton was established.    The collision between two    solitons is found to be quasi elastic at large velocities  with the conservation of total density of each vector soliton.  The component densities are not conserved as the SO-coupled GP equation is not 
Galilean invariant \cite{sol1d}.
 The collision dynamics at small velocity of two vector solitons   is found to be 
inelastic with the destruction of the individual solitons.

  In Sec. \ref{Sec-II}, we describe the mean-field model GP equation
for  a SO-coupled spin-$1$ polar  BEC and  provide an  analytic variational 
solution of this model  using a Gaussian and a hyperbolic secant form of the wave-function  profile to study the SO-coupled spin-1 polar soliton. In 
Sec. \ref{Sec-III}, we provide a numerical solution of the  model by imaginary-time 
propagation and compare the results for density and energy with the corresponding 
analytic variational results. We also study the dynamics of the vector soliton by real-time 
propagation. The   solitons were demonstrated to be dynamically stable. 
The collision dynamics of two  vector solitons was also studied at different 
colliding velocities.
  In Sec. \ref{Sec-IV}  a summary of our findings is presented.


\section{Mean-field model for a SO-coupled BEC}
\label{Sec-II}


We consider a SO-coupled spinor BEC  confined in 1D  along the $x$ axis.  
The 1D confinement  is realized by  strong traps in $y$ and $z$ directions, so that 
the essential 
dynamics  of the system  takes place along  the $x$ axis \cite{Salasnich}, 
while in the transverse $y$ and $z$ directions 
the system is frozen in Gaussian ground states.  
The single particle Hamiltonian of the 
condensate  in  this  quasi-1D trap is  taken  in scaled dimensionless units $\hbar={\bar {\bf m}}=1$ as \cite{usedspin1/2} 
\begin{equation}
H_0 = \frac{p_x^2}{2} + \gamma p_x \Sigma_y,
\label{sph} 
\end{equation}
where $\bar {\bf m}$ is the mass of an atom, $p_x = -i\partial/\partial x$ is the momentum operator along $x$
axis,  and $\Sigma_y$ is the irreducible representation of the $y$ component of 
the spin-1 matrix:  
\begin{eqnarray}
\Sigma_y= \frac{i}{\sqrt 2} \left( \begin{array}
 {ccccc}
0 & -1 & 0\\
1 & 0 & -1\\
0 & 1 & 0 \end{array} \right).
\end{eqnarray}
As we will be investigating vector solitons,  we will not include any trapping potential 
in the Hamiltonian. 

Using the single particle model Hamiltonian  (\ref{sph}) and considering 
interactions in the  Hartree approximation, the  1D \cite{Salasnich} spin-1 BEC of $N$ atoms
can be described by the following set of three coupled mean-field partial 
differential time-dependent  GP equations for the wave-function components $\phi_j$ 
\cite{Ohmi}
\begin{align}
 i \frac{\partial \phi_{\pm 1}}{\partial {t}} &=
 \left( -\frac{1}{2}\frac{\partial^2}{\partial  {x}^2}
 +   {c}_0 {\rho}\right)\phi_{\pm 1} {\mp}  \frac{ { \gamma}}{\sqrt 2}
  \frac{\partial\phi_{0}}{\partial   x}\nonumber\\ 
 &+c_2( {\rho}_{\pm 1}+ {\rho}_0- {\rho}_{\mp 1})\phi_{\pm 1}+ c_2 \phi_0^2\phi_{\mp 1}^*,\label{gp1}\\
 i\frac{\partial \phi_0}{\partial {t}} &= 
 \left( -\frac{1}{2}\frac{\partial^2}{\partial  {x}^2}
 + {c}_0 {\rho}\right)\phi_0  {+}    \frac{ { \gamma}}{\sqrt 2}\left[  \frac{\partial\phi_{+1}}{\partial   x}
 {-}    \frac{\partial\phi_{-1}}{\partial   x}
 \right]
 \nonumber\\
  &  
  +c_2( {\rho}_{+1}+ {\rho}_{-1})\phi_0+ 2 {c}_2\phi_0^*\phi_{+1}\phi_{-1}, \label{gp3}
\end{align} 
where  we have suppressed the   space and time dependence of the wave function $\phi_{\pm 1,0}(x,t) $, $c_0$ and $c_2$ are 
the interaction strengths, $ {\rho}_j = |\phi_j|^2$ where $j=+1,0,-1$ are the  component densities
 corresponding to the three spin components, 
and  ${\rho} = (\rho_{+1}+\rho_0+\rho_{-1})$ is the conserved total density 
   normalized to unity, i.e., 
$
 \int_{-\infty}^{\infty} {\rho}( {x})d {x} = 1. 
$ The conserved magnetization is defined as $\int _{-\infty}^{\infty}
dx  (\rho_{+1}- \rho_{-1}) =m.$

The energy functional corresponding to  the mean-field  SO-coupled spinor BEC model 
(\ref{gp1}) and (\ref{gp3}) is
\cite{Bao}
\begin{align} \label{energy}
 E(\gamma) &\equiv  E_1(\gamma) + E_2(\gamma)\nonumber\\
&= \int_{-\infty}^{\infty} dx \bigg[\frac{1}{2}\left|\frac{d\phi_{+1}}{dx}
  \right|^2+\frac{1}{2}\left|\frac{d\phi_0}{dx}\right|^2+
  \frac{1}{2}\left|\frac{d\phi_{-1}}{dx}\right|^2    \nonumber\\
  &+  \frac{c_0}{2}\rho^2 +  \frac{c_2}{2}\Big\{ \rho_{+1}^2 +  \rho_{-1}^2 + 2
  \big(\rho_{+1}\rho_0 - \rho_{-1} 
  \rho_{+1}\nonumber\\
   &+  \rho_{-1} \rho_0
+
  \phi_{-1}^*\phi_0^2\phi_{+1}^*+\phi_{-1}\phi_{0}^{*2}\phi_{+1}\big)  \Big\} \bigg]
\nonumber \\
& + \int_{-\infty}^{\infty} dx \bigg[
    \frac{ \gamma}{\sqrt 2}\Big\{ \phi_0^* \Big( \frac{d\phi_{+1}}{dx} 
  - \frac{d\phi_{-1}}{dx}\Big)
\nonumber \\&
- \left(\phi_{+1}^*- \phi_{-1}^*\right)\frac{d\phi_0}{dx} 
\Big\}  \bigg].
\end{align}
For magnetization $m=0$, it is possible to develop a variational approximation 
for the stationary problem valid for a polar BEC ($c_2>0$). For $m=0$, 
 we minimize this energy functional using an analytic variational wave function 
to find the analytic solution of the SO-coupled GP equation.

A stationary vector soliton is described by the time-independent version of GP equations  (\ref{gp1}) and (\ref{gp3}) obtained by replacing the terms $i\partial \phi_j/dt$ by $\mu_j\phi_j$, where $\mu_j$s are the chemical potentials.
An analytic approximation scheme for the solution of these time-independent equations 
for a stationary vector soliton
is constructed by  approximating  these solutions in terms of a linear combination of   analytic solutions of the linear version of these equations obtained by setting $c_0=0$, and  $c_2=0$:
\begin{align}
  { E \phi_{\pm 1}}  &=
  -\frac{1}{2}\frac{\partial^2\phi_{\pm 1}}{\partial  {x}^2}
  {\mp}  \frac{ { \gamma}}{\sqrt 2}
  \frac{\partial\phi_{0}}{\partial   x} ,\label{GP1}\\
 { E \phi_0}  &= 
  -\frac{1}{2}\frac{\partial^2\phi_0 }{\partial  {x}^2}
   {+}    \frac{ { \gamma}}{\sqrt 2}\left[  \frac{\partial\phi_{+1}}{\partial   x}
 {-}    \frac{\partial\phi_{-1}}{\partial   x}
 \right]
 , \label{GP3}
\end{align} 
with $E$ the energy. Equations (\ref{GP1}) and (\ref{GP3}) have the following 
degenerate solutions  
\begin{align}\label{s1}
\widetilde \Phi_{1}(x)\equiv 
\left( \begin{array}{c} \widetilde\phi_{+1}\\
\widetilde\phi_0\\
\widetilde\phi_{-1}
 \end{array} \right)
= \frac{e^{ i\gamma x}}{ 2}\left( \begin{array}{c}
1 \\
-i\sqrt{2} \\
-1 \end{array} \right), \\  \widetilde  \Phi_2  (x)  \equiv 
\left( \begin{array}{c} \widetilde\phi_{+1}\\
\widetilde\phi_0\\
\widetilde\phi_{-1}
 \end{array} \right)
= \frac{e^{ -i\gamma x}}{ 2}\left( \begin{array}{c}
1 \\
i\sqrt{2} \\
-1 \end{array} \right), \label{s2}
\end{align}
each of  energy $E=-\gamma^2 /2$.

 The analytic ansatz for the stationary wave functions of Eqs. (\ref{gp1}) and (\ref{gp3}) are taken as  the following  linear combination of solutions (\ref{s1}) and (\ref{s2}): 
\begin{align}
\Phi_1(x)\equiv  \left( \begin{array}{c} \phi_{+1}\\
\phi_0\\
\phi_{-1}
 \end{array} \right)=&  \frac{\psi (x)}{\sqrt{2}}  \left[\widetilde \Phi_{1}(x) + \widetilde \Phi_2(x)\right]  \nonumber \\
=& \frac{\psi(x)}{\sqrt 2}\left( \begin{array}{c}
\cos(\gamma x) \\
\sqrt{2}\sin(\gamma x) \\
-\cos(\gamma x) \end{array} \right), \label{aussian}\\
\Phi_2(x)\equiv  \left( \begin{array}{c} \phi_{+1}\\
\phi_0\\
\phi_{-1}
 \end{array} \right)=&  \frac{\psi (x)}{\sqrt{2}}  \left[\widetilde \Phi_{1}(x) - \widetilde \Phi_2(x)\right]  \nonumber \\  
=& \frac{i\psi(x)}{\sqrt 2}\left( \begin{array}{c}
\sin(\gamma x) \\
-\sqrt{2}\cos(\gamma x) \\
-\sin(\gamma x) \end{array} \right), \quad
\label{aussian2}
\end{align}
where $\psi(x)$ is a normalized localized  function. 
 The forms (\ref{aussian})
and (\ref{aussian2}) are also motivated by  a prior study of numerical results,  which reveals that the component densities have the same  localized shape \cite{abc}, but with different normalizations,  modulated  sinusoidally with a 
spatial frequency of  $\gamma/\pi$ and that a maximum density of components  $j = \pm 1$ is accompanied by a minimum density of component $j=0$ and vice versa.   
The spatial density modulation frequency   $\gamma/\pi$ appears naturally by construction through the sine and cosine terms.

 We will consider  the following Gaussian  and 
secant hyperbolic ansatz for  the  function $\psi(x)$ in Eqs. (\ref{aussian})-(\ref{aussian2})
\begin{align}\label{gauss}
\psi(x)& = {\frac{1}{\sqrt{\alpha\sqrt \pi }}}\exp\left[-\frac{x^2}{2\alpha^2}\right], \\
\psi(x) &= \frac{\sqrt{\sigma}}{\sqrt 2}  {\mathrm{sech}}(\sigma x).\label{sech}
\end{align}
With these ansatz for the wave function, the energy functional (\ref{energy}) is explicitly real,  has the correct $\gamma$ dependence,   magnetization ($m=0$) and normalization ($=1$).   
Ansatz (\ref{aussian}) for the wave function  has  maxima of density 
at center ($x=0$) in components $j=\pm 1$   given by  cosine functions and a minimum (zero) in density at center in component $j=0$
 and is a
vector soliton of type bright-dark-bright.  Ansatz (\ref{aussian2}), on the other hand,   corresponds to a maximum of density  at center  in component $j=0$ 
accompanied by minima (zero) of density at center in components $j=\pm 1$     and is a vector soliton of type dark-bright-dark.  The numerical solution has exactly the same behavior.  As confirmed by our numerical solution and analytic approximation, these two types of vector solitons 
have the same energy and are degenerate states.

With the analytic ansatz (\ref{aussian}), and also (\ref{aussian2}), for the profile of the vector soliton,  the energy functional
 (\ref{energy}) with function (\ref{gauss})  can be evaluated to yield
\begin{align}\label{en1}
E(\gamma)=-\frac{\gamma^2}{2}+\frac{1}{4\alpha^2}  +  \frac{c_0}{2 \alpha \sqrt{2\pi}}.   
\end{align} 
The width $\alpha$ of the minimum-energy ground state vector soliton  is obtained by minimizing this energy functional with respect to $\alpha$:  
\begin{align}\label{minal}
\alpha = -\frac{\sqrt{2\pi}}{c_0}.
\end{align}
For this width to be positive we 
require $c_0 <0$.   This width  is independent of the SO-coupling  strength  $\gamma$ and interaction strength $c_2$.  The following  minimum of energy as
a function of $\gamma$ is  obtained by substituting Eq. (\ref{minal}) in Eq. (\ref{en1})
\begin{align}\label{En1}
E(\gamma) = -\frac{\gamma^2}{2} -  \frac{c_0^2}{8\pi} ,
\end{align}
which is the energy of the minimum-energy spin-1 three-component vector soliton in the ground state. The corresponding expression 
for the function (\ref{gauss})  is
\begin{align}\label{auss}
\psi(x) = \left(\frac{c_0^2}{2\pi^2}   \right)^{1/4}   \exp \left( - \frac{c_0^2 x^2}{4 \pi}    \right) .
\end{align}
 We note that the analytic variational results  (\ref{En1})  and (\ref{auss}) for the energy and the wave function, respectively, are determined by the interaction parameter  $c_0$ only and independent of the
 interaction strength $c_2$.  

With both ansatz (\ref{aussian}) and (\ref{aussian2}), employing the secant hyperbolic function (\ref{sech}), the energy 
functional becomes \cite{sol1d}
\begin{align}\label{en2}
E(\gamma)=-\frac{\gamma^2}{2}+ \frac{\sigma^2 +c_0\sigma}{6},   
\end{align}
which is minimized at 
\begin{align}
\sigma =-\frac{c_0}{2}
\end{align}
yielding the energy minimum
\begin{align}\label{En2}
E(\gamma)=-\frac{\gamma^2}{2}-\frac{c_0^2}{24}
\end{align}
and the function (\ref{sech})
\begin{align} \label{ech}
\psi(x)=\frac{\sqrt{|c_0|}}{2}\mathrm{sech}\left(\frac{|c_0|x}{2}\right).
\end{align}
We find that the analytic energy (\ref{En2})  is smaller than  energy (\ref{En1}).
Due to the variational nature of the analytic calculation,  energy (\ref{En2}) is closer to the 
exact energy and hence the secant hyperbolic form (\ref{ech}) of the wave function is a better 
approximation  to the exact solution
 than the Gaussian form (\ref{auss}).

 In one dimension, for a spin-1 spinor BEC,  there are two other linearly independent SO couplings:  $\gamma p_x \Sigma_x$ \cite{sol1d} and  $\gamma p_x \Sigma_z$ \cite{usedspin1/2}, where 
\begin{align} 
\Sigma_x= \frac{1}{\sqrt{2}}\left( \begin{array}
 {ccccc}
0 & 1 & 0\\
1 & 0 & 1\\
0 & 1 & 0 \end{array} \right),\quad 
\Sigma_z= \left( \begin{array}
 {ccccc}
1 & 0 & 0\\
0 & 0 & 0\\
0 & 0 & 1 \end{array} \right),
\end{align}
Of these, the SO coupling    $\gamma p_x \Sigma_x$, like the  SO coupling $\gamma p_x \Sigma_y$, 
 connects the components $F_z=\pm 1$ with the component $F_z=0$  
and these two SO couplings are equivalent and have identical analytic variational solution.
 To demonstrate  this claim explicitly,  we note that for SO coupling   $\gamma p_x \Sigma_x$
 the mean-field GP equation is  \cite{gautam-2}
\begin{align}
 i \frac{\partial \phi_{\pm 1}}{\partial {t}} &=
 \left( -\frac{1}{2}\frac{\partial^2}{\partial  {x}^2}
 +   {c}_0 {\rho}\right)\phi_{\pm 1} {\color{red}-}  \frac{ {i \gamma}}{\sqrt 2}
  \frac{\partial\phi_{0}}{\partial   x}\nonumber\\ 
 &+c_2( {\rho}_{\pm 1}+ {\rho}_0- {\rho}_{\mp 1})\phi_{\pm 1}+ c_2 \phi_0^2\phi_{\mp 1}^*,\label{gps-1}\\
 i\frac{\partial \phi_0}{\partial {t}} &= 
 \left( -\frac{1}{2}\frac{\partial^2}{\partial  {x}^2}
 + {c}_0 {\rho}\right)\phi_0  {\color{red}-}    \frac{ {i \gamma}}{\sqrt 2}\left[  \frac{\partial\phi_{+1}}{\partial   x}
 {\color{red}+}    \frac{\partial\phi_{-1}}{\partial   x}
 \right]
 \nonumber\\
  &  
  +c_2( {\rho}_{+1}+ {\rho}_{-1})\phi_0+ 2 {c}_2\phi_0^*\phi_{+1}\phi_{-1}.\label{gps-3}
\end{align}
The energy functional is again given by (\ref{energy})  but now with $E_2(\gamma)$ given by 
\begin{align}
E_2(\gamma)&= 
\int_{-\infty}^{\infty} dx \bigg[
   - \frac{i \gamma}{\sqrt 2}\Big\{ \phi_0^* \Big( \frac{d\phi_{+1}}{dx} 
  + \frac{d\phi_{-1}}{dx}\Big)
\nonumber \\&
+ \left(\phi_{+1}^*+ \phi_{-1}^*\right)\frac{d\phi_0}{dx} 
\Big\}  \bigg].
\end{align} 

The analytic ansatz for the wave function in this case can be one of the following forms 
\begin{align}
\Phi_1
\equiv & \left( \begin{array}{c} \phi_{+1}\\
\phi_0\\
\phi_{-1}
 \end{array} \right)
= \frac{\psi(x)}{\sqrt 2}\left( \begin{array}{c}
\cos(\gamma x) \\
-i\sqrt{2}\sin(\gamma x) \\
\cos(\gamma x) \end{array} \right), \quad
\label{Gaussian}\\
\Phi_2\equiv & \left( \begin{array}{c} \phi_{+1}\\
\phi_0\\
\phi_{-1}
 \end{array} \right)
= \frac{\psi(x)}{\sqrt 2}\left( \begin{array}{c}
\sin(\gamma x) \\
i\sqrt{2}\cos(\gamma x) \\
\sin(\gamma x) \end{array} \right). \label{Gaussian2}
\end{align}
With these forms the energy functional are explicitly real. Although the wave functions 
(\ref{Gaussian}) and (\ref{Gaussian2}) on one hand, and (\ref{aussian}) and (\ref{aussian2})
on the other hand are different, they lead to  the same densities $\rho_j(x)$ and energies $E(\gamma)$. With the forms 
 (\ref{Gaussian}) and (\ref{Gaussian2}), the expression for the energy functional for a BEC 
with SO coupling  $\gamma p_x \Sigma_x$ becomes identical with the expression for the energy 
functional for a BEC 
with SO coupling  $\gamma p_x \Sigma_y$  for both Gaussian and secant hyperbolic ansatz 
(\ref{gauss})  and (\ref{sech}) for the function $\psi(x)$. Hence the analytic solutions for the 
function $\psi(x)$ for SO coupling $\gamma p_x \Sigma_x$  are also given  by (\ref{auss}) and 
(\ref{ech}), respectively, for the Gaussian and secant hyperbolic functions.  However, 
the SO  coupling $\gamma p_x \Sigma_z$  has a distinct coupling scheme and will be considered elsewhere.

\section{Result and Discussion}
\label{Sec-III}

We numerically solve the coupled partial differential equations  (\ref{gp1})-(\ref{gp3}) using the
split-time-step Crank-Nicolson method \cite{Muruganandam} by real- and imaginary-time propagation methods.
For a numerical simulation there are the Fortran \cite{Muruganandam} and C  \cite{cc} programs  and their 
open-multiprocessing \cite{omp} versions  appropriate for using in multi-core processors. 
 The ground 
state is determined by solving  (\ref{gp1})-(\ref{gp3}) using 
imaginary time propagation, which neither conserves normalization  nor magnetization. 
Both normalization  and magnetization can be fixed by renormalizing  the 
wave-function components appropriately after each time iteration \cite{Bao}.  
The initial wave function in imaginary-time propagation is taken as the variational function (\ref{aussian})
or  (\ref{aussian2})  
with the Gaussian or secant hyperbolic form for the function $\psi(x)$ given by (\ref{auss}) or (\ref{ech}), respectively. The choice (\ref{aussian}) leads to a vector soliton of type 
bright-dark-bright and the choice (\ref{aussian}) generating the degenerate  one of type dark-bright-dark.  If we just take a localized function as the initial state without sine or
cosine modulation the final state will be one of these two degenerate solutions. 
The real-time propagation method was used to study the dynamics with the converged solution     
obtained in imaginary-time propagation as the initial state. 
The space and time steps employed in the imaginary-time propagation are $dx =0.025$ and $dt =0.0001$ 
and that in the real-time propagation are  $dx =0.025$ and $dt =0.00001$.

\begin{figure}[!t]
\begin{center}
\includegraphics[trim = 0mm 0mm 0cm 0mm, clip,height=4cm,width= 7.5cm,clip]{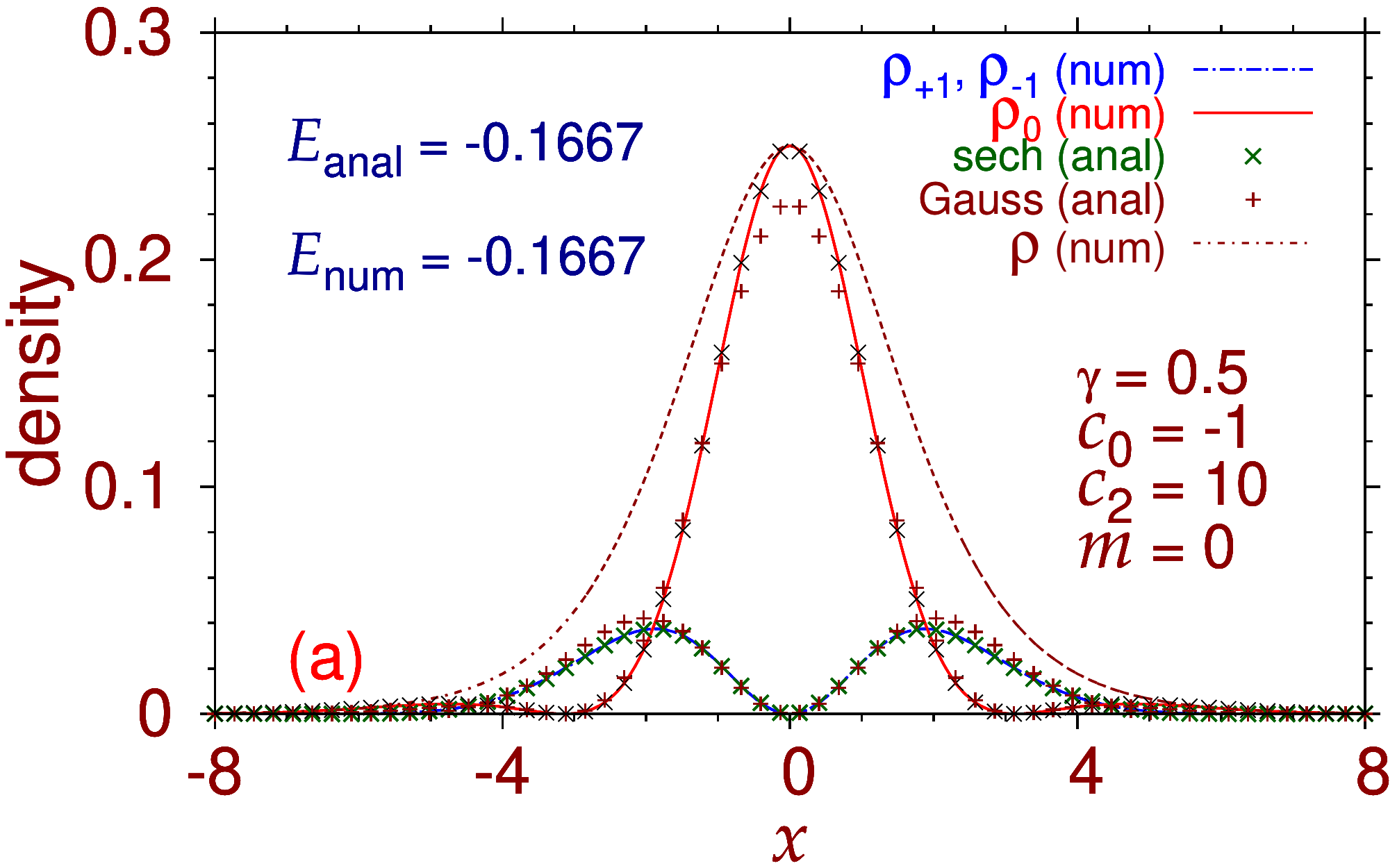}
\includegraphics[trim = 0mm 0mm 0cm 0mm, clip,height=4cm,width= 7.5cm,clip]{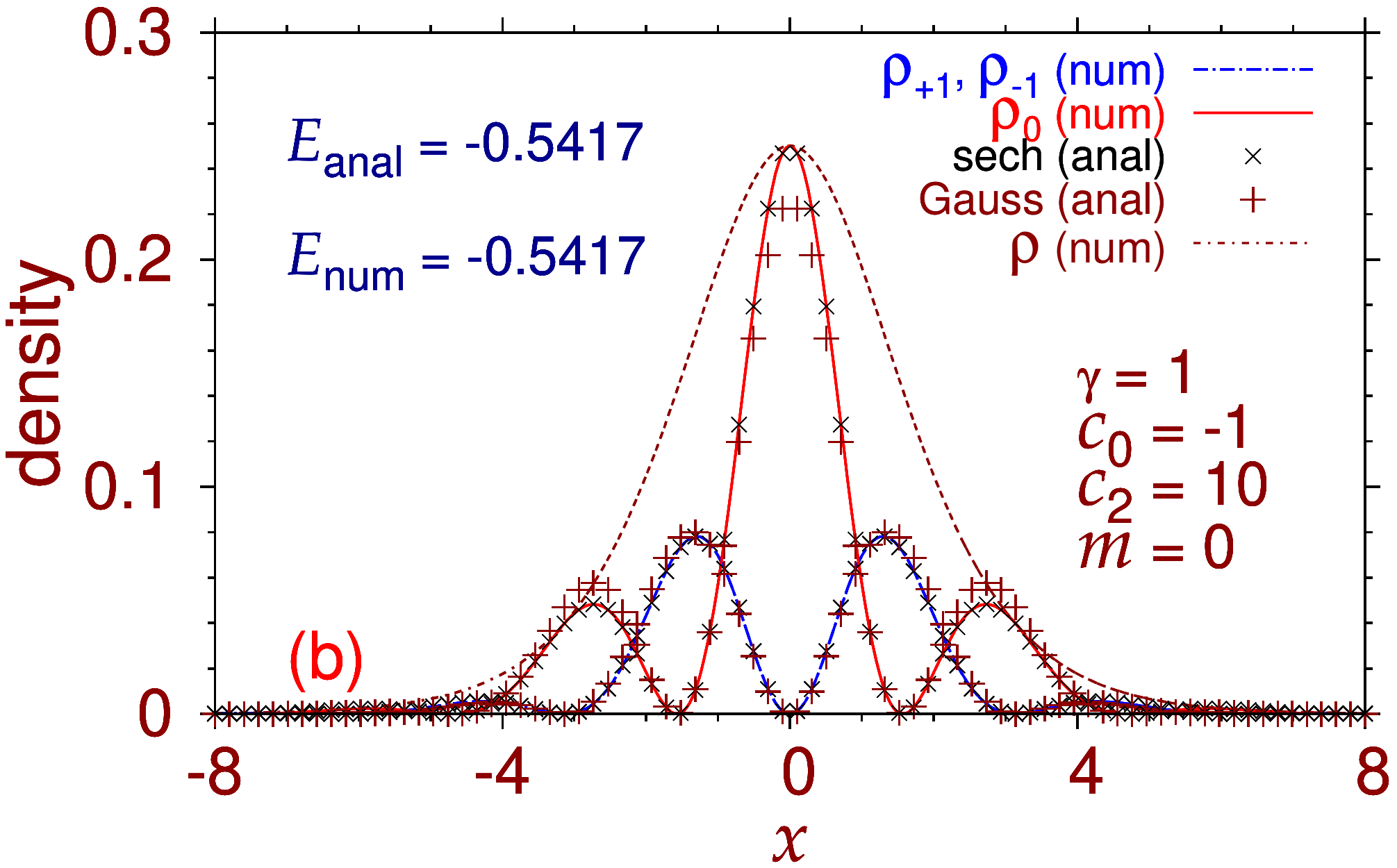} 
\includegraphics[trim = 0mm 0mm 0cm 0mm, clip,height=4cm,width= 7.5cm,clip]{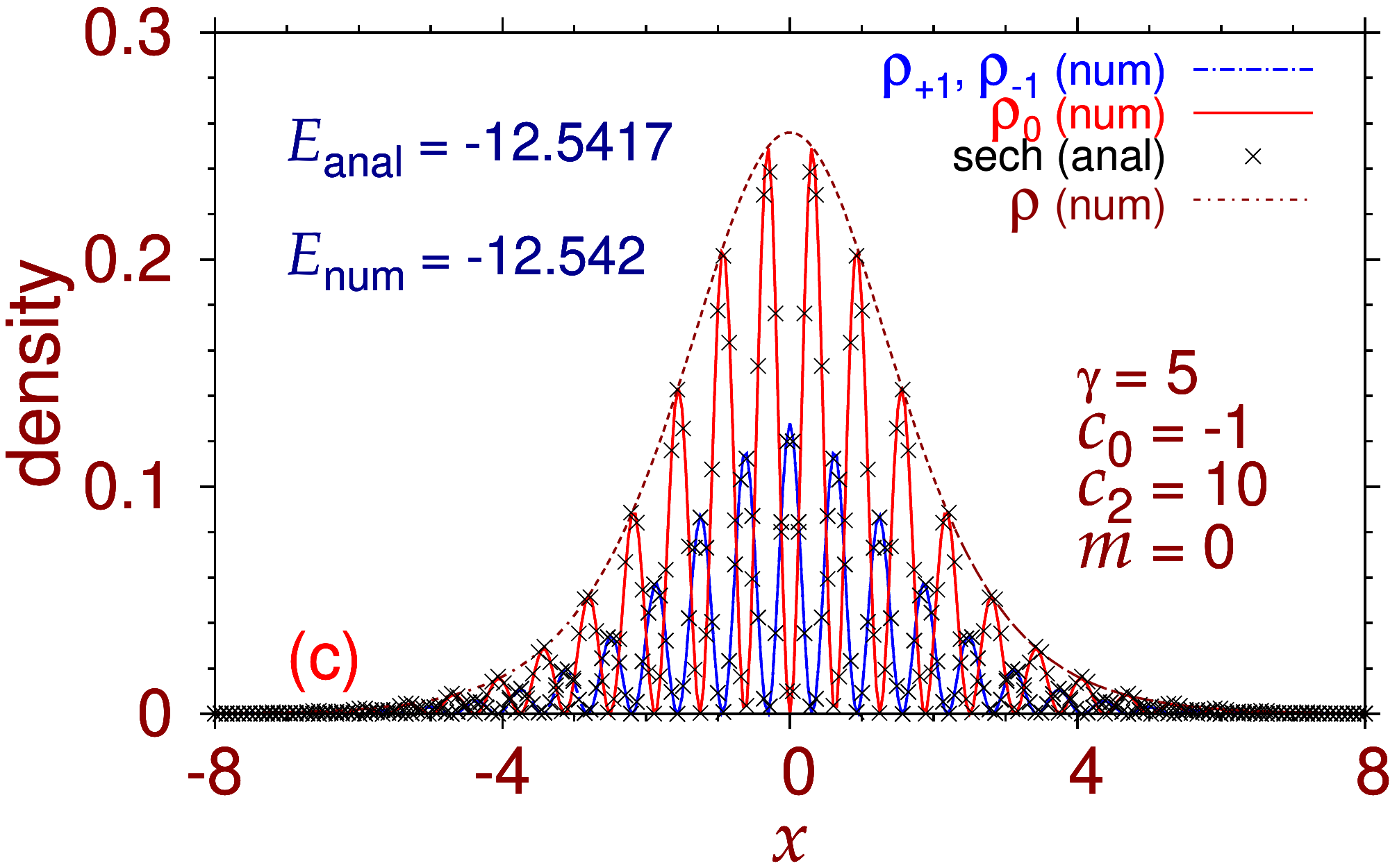}

\caption{(Color online)   Numerical (num) and analytic variational (anal) densities with the secant hyperbolic (sech) and Gaussian (Gauss) functions  $\rho_j(x) \equiv |\phi_j|^2, j=0, \pm 1$ 
of the three components for the lowest-energy    vector  soliton 
 with $c_0 = -1$, $c_2 = 10, m=0$,  for (a) $\gamma =0.5,$  (b)    $\gamma =1$, and (c) $\gamma =5$.   
All quantities in this and following figures  are dimensionless.}
\label{fig1} \end{center}
\end{figure}

\begin{figure}[!t]
\begin{center}
 
\includegraphics[trim = 0mm 0mm 0cm 0mm, clip,height=4cm,width= 7.5cm,clip]{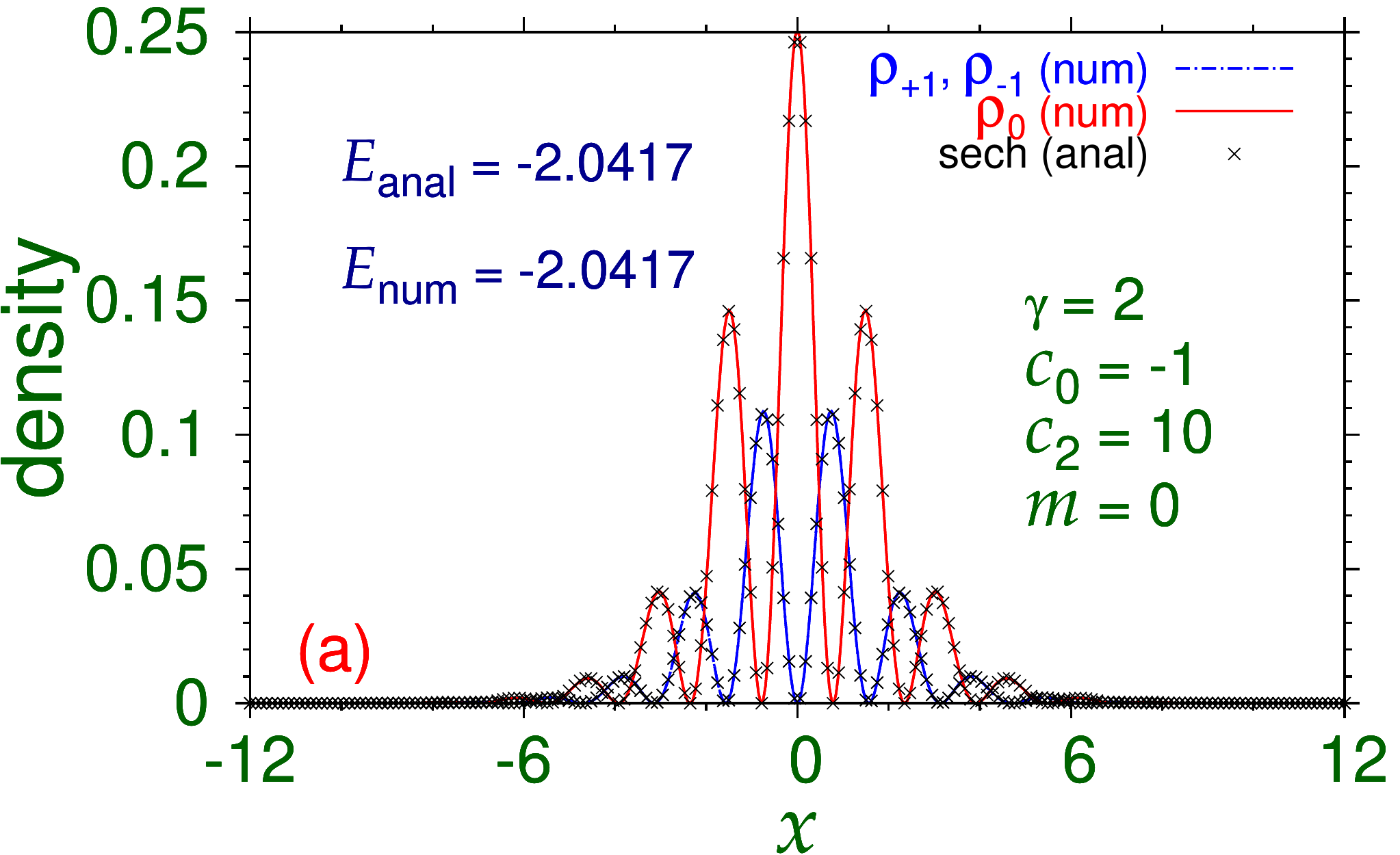}
\includegraphics[trim = 0mm 0mm 0cm 0mm, clip,height=4cm,width= 7.5cm,clip]{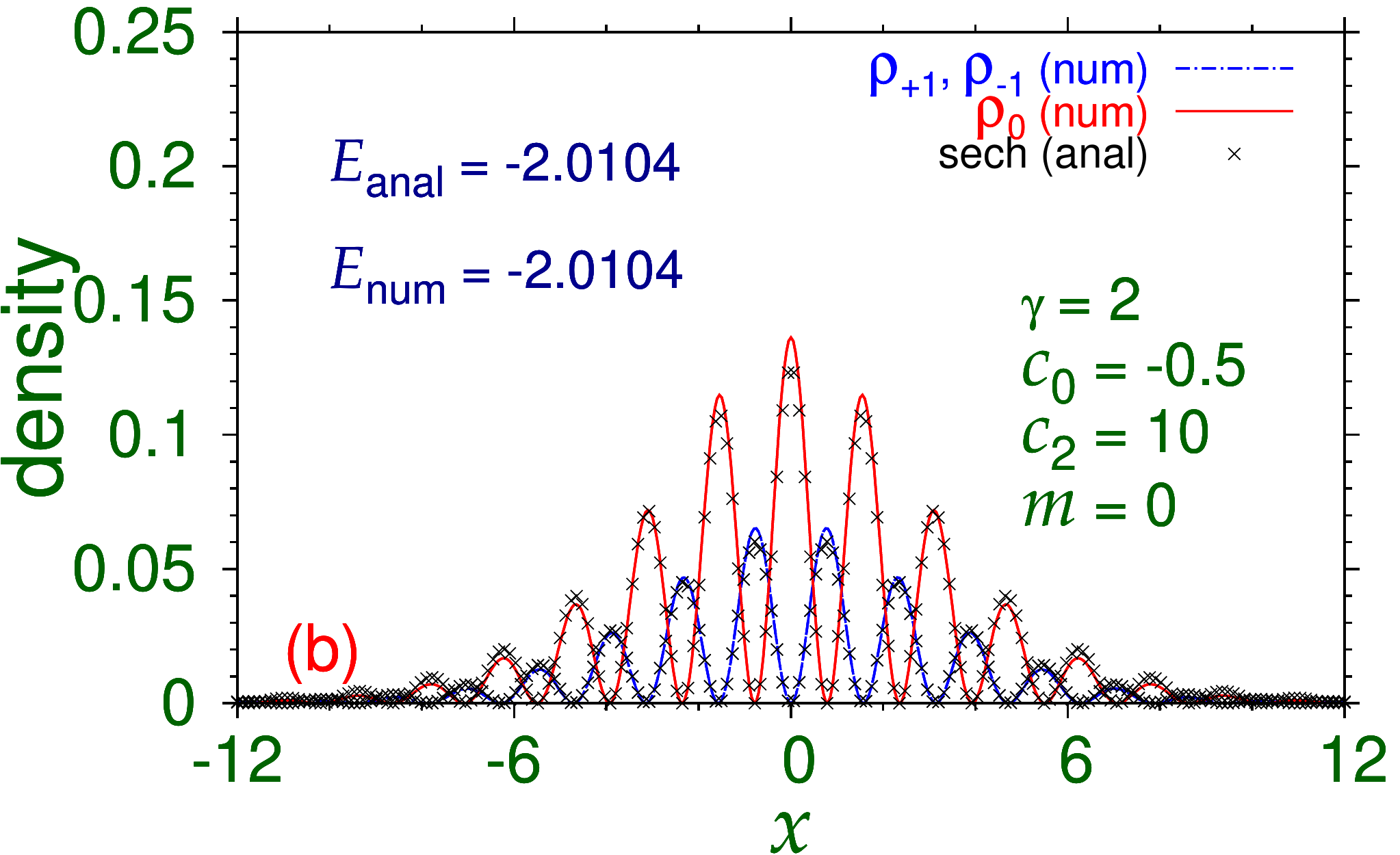} 

\caption{(Color online)   Numerical (num) and analytic variational (anal) densities with the secant hyperbolic (sech)   function
  $\rho_j(x) \equiv |\phi_j|^2, j=0, \pm 1$ 
of the three components for the lowest-energy    vector  soliton 
 with $\gamma  = 2$, $c_2 = 10, m=0$,  for (a) $c_0 =-1,$  (b)    $c_0 =-0.5$. 
      }

\label{fig2} \end{center}

\end{figure}

We begin our calculation with parameters $c_0=-1$   and $c_2= 10$ ($>0$, polar),
to make the system attractive to have a vector soliton of a reasonable width (not too large or small) with magnetization $m=0$. A larger value of 
$|c_0|$  will increase attraction and reduce the width and vice versa, viz. Eq. (\ref{minal}).
 In Figs. \ref{fig1}(a)-(c) we display the density of the components of the minimum-energy   ground-state vector soliton for $\gamma = 0.5, 1, $  and 5, respectively. The result  of the analytic variational approximation with the hyperbolic secant function (\ref{ech})  is also displayed in these plots.  In Fig. \ref{fig1}(a) and (b)   we also show the analytic result with the Gaussian function (\ref{auss}).  In these plots we note, as expected, the analytic approximation with the secant hyperbolic function is better than the that  with the Gaussian function. Hence we will show in the following only the analytic result using the secant hyperbolic function.     The variational result for the width $\alpha$ given by (\ref{minal}) is the same for all components. The same is found to be true in the numerical calculation, in good agreement with the analytic approximation.  Apart from the component densities $\rho_j(x)$, we also exhibit in these plots the total numerical density $\rho(x) =\sum_j \rho_j(x) $ in excellent agreement with the analytic result obtained with the secant hyperbolic function (not shown in these plots). The component densities have pronounced maxima and minima controlled by the functions 
$\cos^2 (\gamma x) $ and $\sin^2 (\gamma x)$, viz. Eqs. (\ref{aussian}) or (\ref{aussian2}). In Fig. \ref{fig1}(c) 14 such maxima are noted.
However, the total density $\rho(x)$ has a localized secant hyperbolic (or Gaussian) shape without any maximum or minimum. 
In Figs. \ref{fig1}(a)  the $j=0$ component has a maximum at the center like a bright soliton and the $j=\pm 1$ components have a minimum at the center like a dark soliton. Hence the vector soliton in this plot is of the type dark-bright-dark.  The same is true in Fig.  \ref{fig1}(b), although in the latter case an undulating tail in component densities starts to appear 
due to a larger value of the SO coupling strength $\gamma$. In this terminology the vector soliton in Fig. \ref{fig1}(c)
with pronounced undulating tails can be termed  of the type bright-dark-bright   because in this case  central 
maxima in components $j=\pm 1$  is accompanied by a central minimum in component $j=0$.  These two types of vector solitons $-$   
dark-bright-dark and bright-dark-bright $-$ are degenerate  corresponding to the same energy eigenvalue.

\begin{figure}[!t]
\begin{center}
\includegraphics[trim = 0mm 0mm 0cm 0mm, clip,height=4cm,width= 7.5cm,clip]{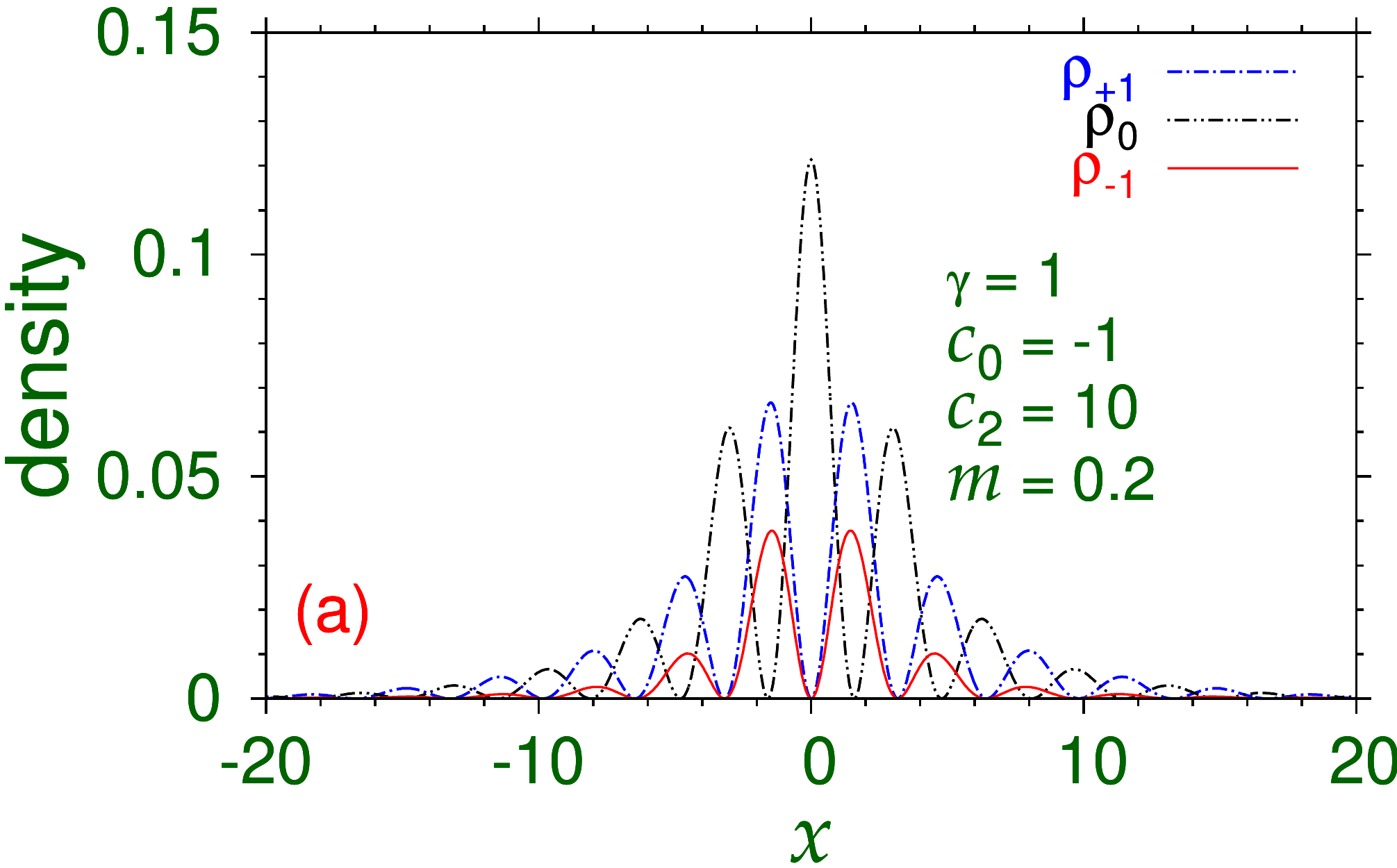}
\includegraphics[trim = 0mm 0mm 0cm 0mm, clip,height=4cm,width= 7.5cm,clip]{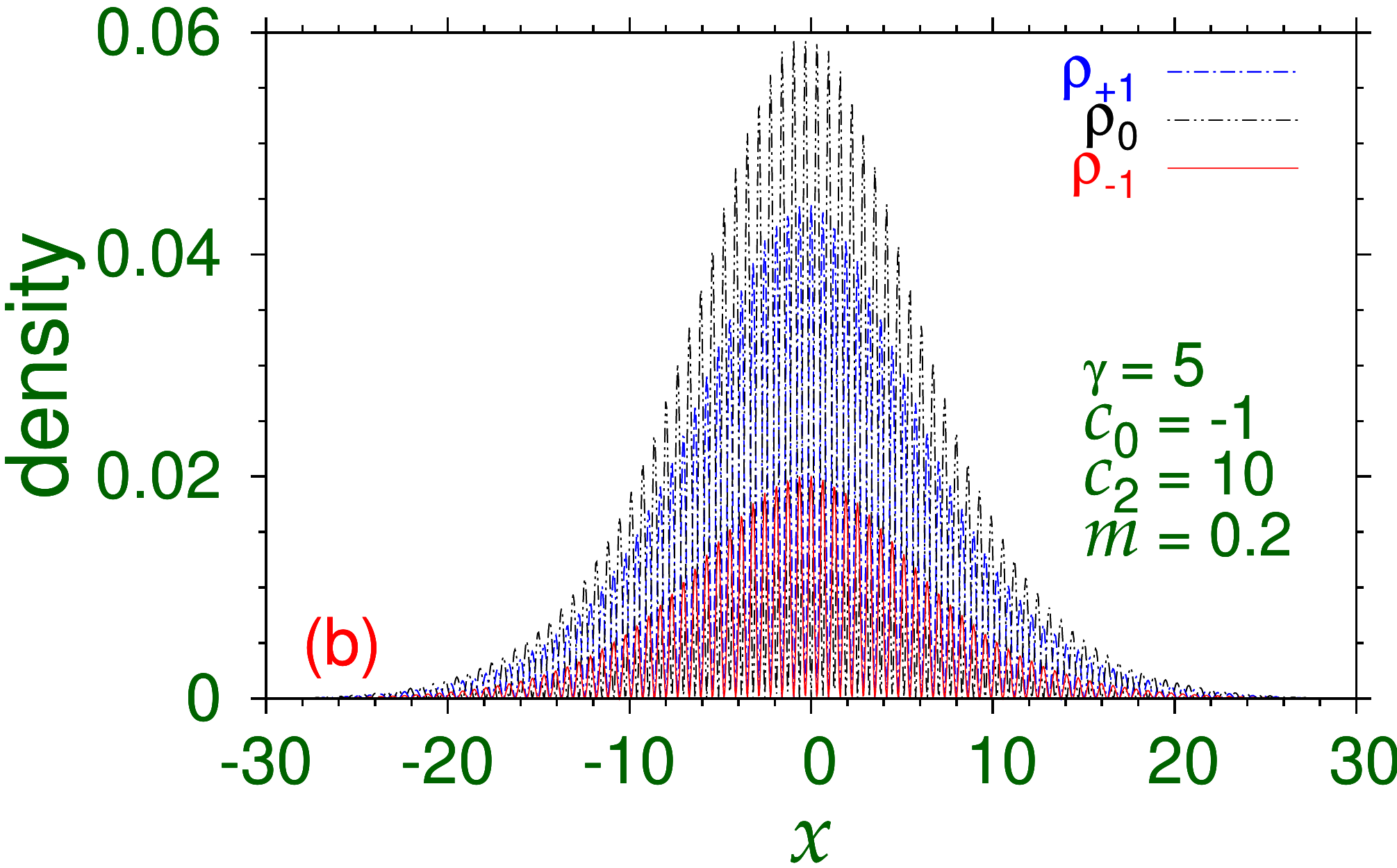}

\caption{(Color online)   Numerical (num)  density
  $\rho_j(x) \equiv |\phi_j|^2, j=0, \pm 1$ 
of the three components for the lowest-energy    vector  soliton 
 with  $c_0=-1, c_2 = 10, m=0.2$,  for (a) $\gamma =1,$  and (b)    $\gamma =5$. 
   }
\label{fig3} \end{center}
\end{figure}

In Figs. \ref{fig1}  we exhibited the pattern of vector solitons for different SO coupling strengths $\gamma$ and find that the number of peaks in the soliton density increases with $\gamma$.  In Figs. \ref{fig2} we show the effect of a variation of the 
interaction strength $c_0$ on the soliton profile. In Figs. \ref{fig2}  we display the soliton profile for $\gamma=2, m=0, c_2=10$ 
and for (a) $c_0 =-1$ and (b) $c_0 =-0.5$. A  reduced value of $|c_0|$ corresponds to reduced attraction resulting in a soliton in 
Fig. \ref{fig1}(b) with larger spatial extension. In  Figs.  \ref{fig1}  we see that a increased value of $\gamma$ leads to an 
increased number of maxima and minima in soliton density due to a larger spatial frequency of undulation. In fact the spatial frequency of density maxima for a wave function profile 
$\cos(\gamma x)$ or $\sin(\gamma x)$   is $\gamma/\pi$. 
In   Figs.  \ref{fig2}  we find that a reduced  value of $|c_0|$ leads to an 
increased number of maxima and minima in soliton density due to a larger spatial extension of the vector soliton.

So far we studied the soliton profiles with zero magnetization  $m=0$.   Next we study these with a  non-zero magnetization $m=0.2$. 
A non-zero magnetization breaks the symmetry between the $j=\pm 1 $ states which now will have different spatial densities. For $m=0$ the densities of the $j=\pm 1 $ states are the same.   In Figs. \ref{fig3} we display only the numerical density profiles with $c_0=-1, c_2=10, m=0.2$ and   (a) $\gamma= 1$   and (b) $\gamma=5$. In this case of non-zero
 $m$, there is no analytic variational result (valid only for $m=0$).  Although the spatial extension of the soliton in Figs. \ref{fig3}(a) and (b) are similar, in the latter case we have many more maxima and minima due to an increased value of $\gamma$ resulting in a larger spatial frequency of density modulation. In Fig. \ref{fig3}(b), for $\gamma =5$, there are more than 60 maxima and minima in density in the spatial range between $x=\pm 20$, the corresponding analytic estimate being $40 \times  \gamma/\pi \sim 63$  in agreement with the numerical result.

\begin{figure}[!t]
\begin{center}
\includegraphics[trim = 0mm 0mm 0cm 0mm, clip,height=4.5cm,width= 7.5cm,clip]{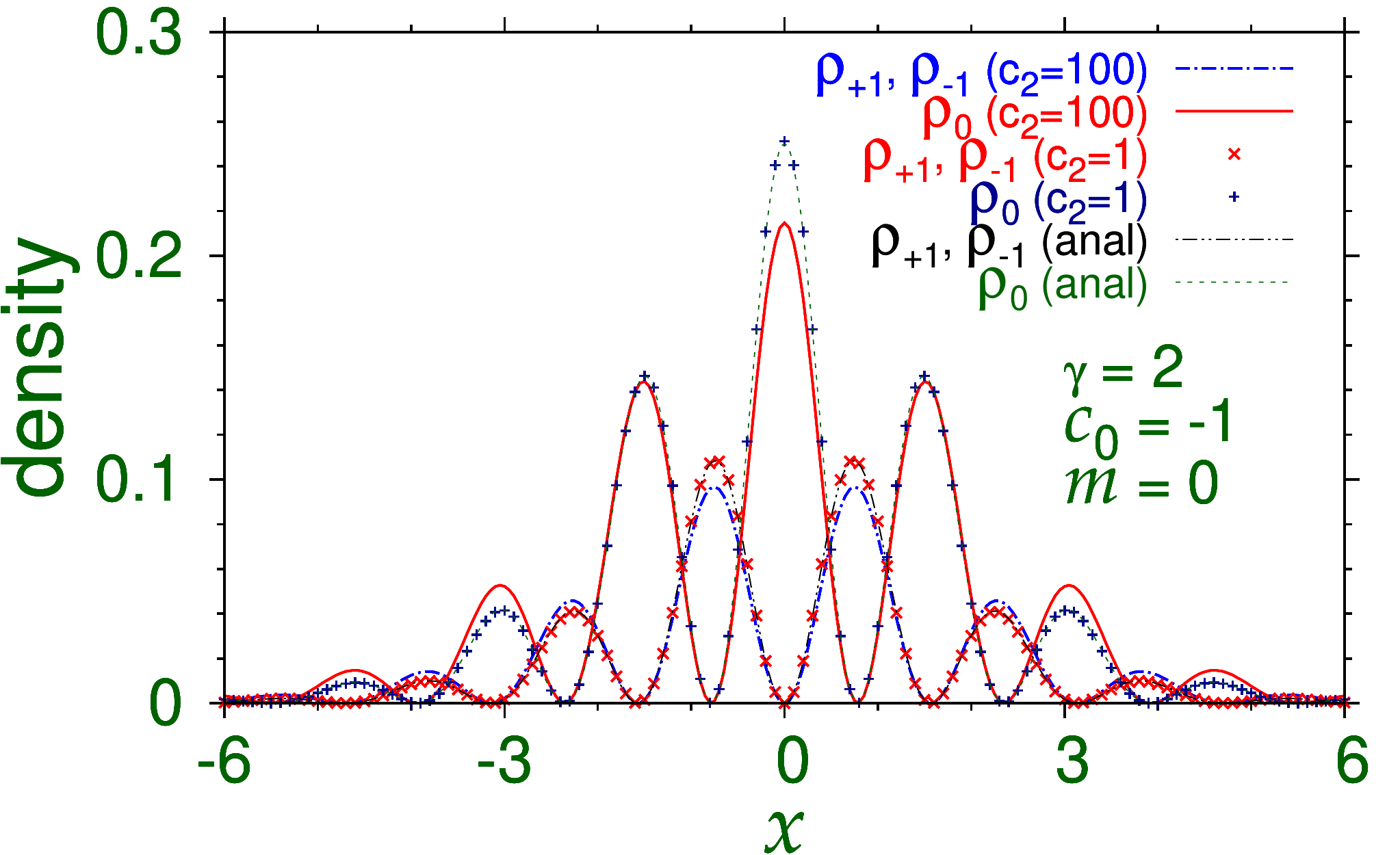} 

\caption{(Color online)    Numerical (num) and analytic variational (anal) densities with the secant hyperbolic    function
  $\rho_j(x) \equiv |\phi_j|^2, j=0, \pm 1$ 
of the three components for the lowest-energy    vector  soliton 
 with $\gamma  = 2$, $c_0 = -1, m=0$,  for $c_2 =1,$     and 100. The analytic densities are in agreement with the results for $c_2=1$.   }
\label{fig4} \end{center}
\end{figure}

 We have seen that the analytic results for density and energy of a vector soliton are independent of the interaction strength $c_2$ being determined 
solely by the strength $c_0$, viz. Eqs. (\ref{En2}) and (\ref{ech}). The numerical results for  density and energy of a vector soliton
are not entirely independent of $c_2$, they have a weak dependence on $c_2$. To demonstrate this, in Fig. \ref{fig4} we plot  the
numerical and analytic  densities of soliton components for $\gamma=2, c_0 = -1, m=0$   and for $c_2=1$  and  100.  The 
analytic densities are in agreement with the numerical result for $c_2=1$  closer to the linear limit where the analytic results are expected  to  be more reliable.

\begin{figure}[!t]
\begin{center}
\includegraphics[trim = 0mm 0mm 0cm 0mm, clip,height=4.5cm,width= 15cm,clip]{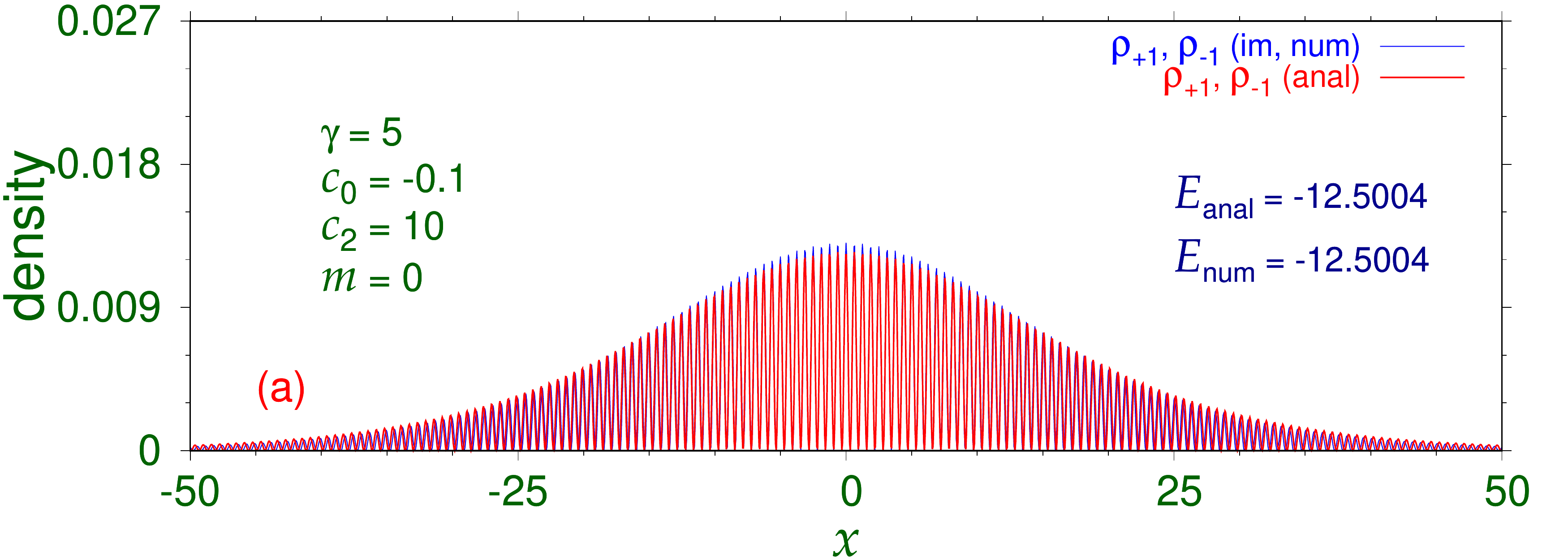}
\includegraphics[trim = 0mm 0mm 0cm 0mm, clip,height=4.5cm,width= 15cm,clip]{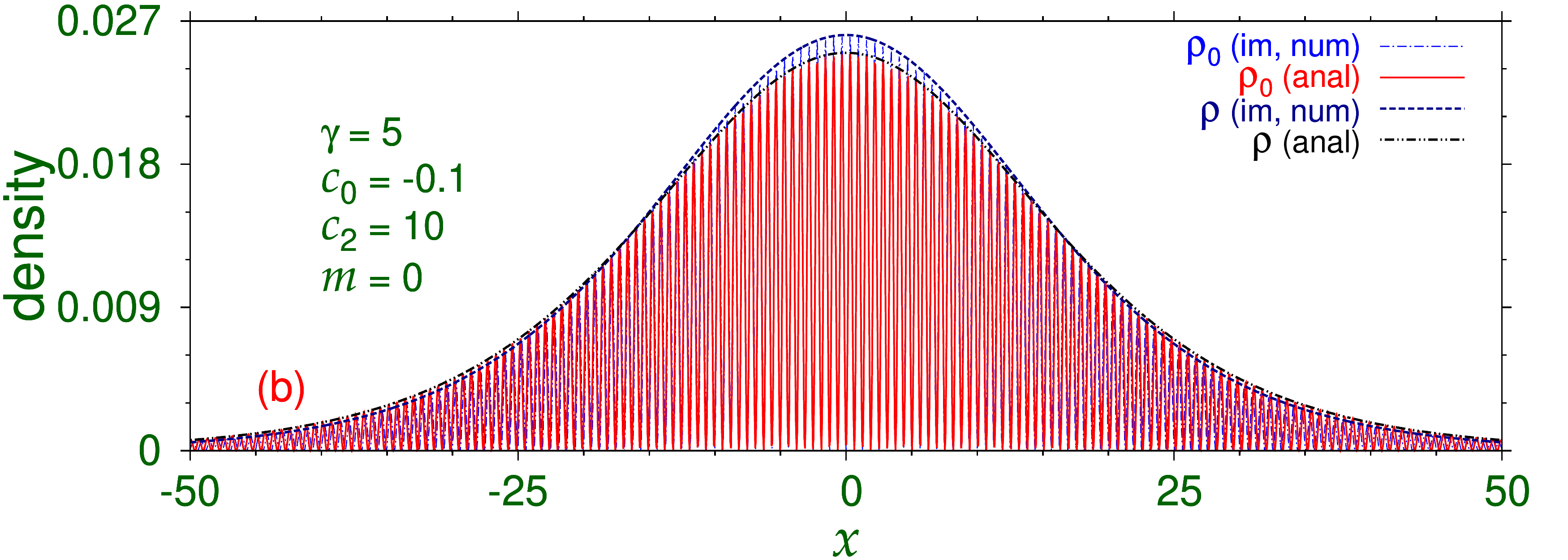} 
\includegraphics[trim = 0mm 0mm 0cm 0mm, clip,height=4.5cm,width= 15cm,clip]{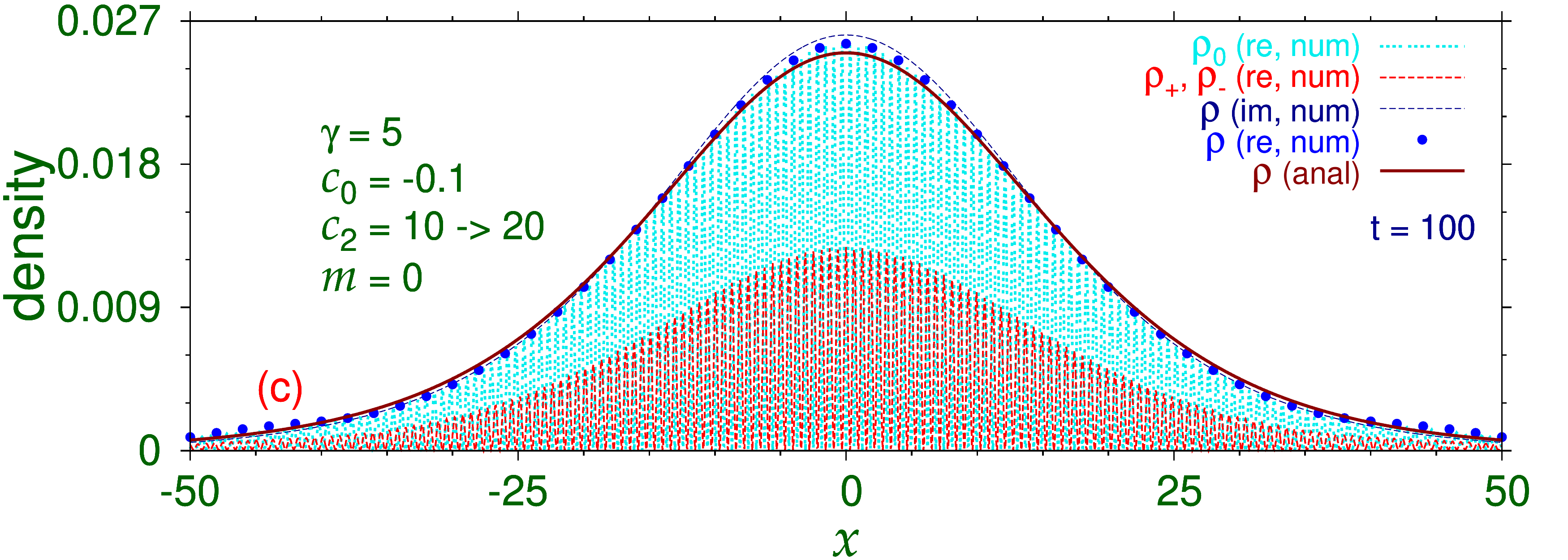}

\caption{(Color online)  (a)-(b) Numerical and analytic variational  densities $\rho_j$, for 
(a) $ j= \pm 1$ and (b)   $j=0$ 
of the three components of  the    vector  soliton 
 with $c_0 = -0.1$, $c_2 = -10$, $\gamma =5, m=0$ by imaginary-time propagation.  In (b) we also display the analytic result 
for $\rho_0$ and $\rho$ 
obtained with the secant hyperbolic function.
(c) Results for component and total densities  after  real-time propagation over an interval of time of 100 units obtained using the converged stationary  state  of  imaginary-time propagation as the initial state after changing $c_2$ from 10 to 20 at time $t=0$.  The results of total density obtained by  imaginary-time propagation  and also by analytic approximation 
are also exhibited. 
 }

\label{fig5} \end{center}

\end{figure}

Now we consider a vector soliton with a very large number of maxima and minima in density with a small $|c_0|$ and large $\gamma$. 
A small $|c_0|$ increases the spatial extension while a large $\gamma$ increases the spatial frequency of density modulation of the components of a stripe vector soliton \cite{stripe}. 
In Figs. \ref{fig5}(a)-(b) we display the soliton densities for components $j=\pm 1$ and $j=0$, respectively, 
calculated with $c_0=-0.1, c_2 =10, m=0, \gamma=5$.  The total soliton density $\rho$ is also displayed in Fig. \ref{fig5}(b). The analytic densities calculated with the secant hyperbolic function are also shown in these plots.  From Figs. \ref{fig5}(a)-(b) we find that the numerical densities are in excellent agreement with their simple analytic approximations. Although the component densities of the vector soliton exhibit a large number of maxima and minima ($> 150$), the total density exhibited in Fig. \ref{fig5}(b), as expected,   has a smooth behavior without any modulation.

To demonstrate that the stripe vector soliton  exhibited in Figs. \ref{fig5}(a)-(b) is dynamically stable, we subject the ground-state vector soliton profile, 
obtained by imaginary-time simulation, to real-time propagation for a long time after giving a perturbation by 
changing the strength of the nonlinear interaction $c_2$ from 10 to 20 at time $t=0$ maintaining other parameters fixed.  The  real-time propagation was executed for  a total time interval of 100.  
     In Fig. \ref{fig5}(c)   we exhibit the density profile of the three 
components  at $t=100$  of the vector soliton at the end of  real-time propagation together with the total density $\rho$.  In this plot we also show the results for total  density $\rho$  obtained in imaginary-time propagation as well as  using the analytic approximation with the secant hyperbolic function.
The 
long-time stable propagation of the components of the vector soliton as shown in Fig. \ref{fig5}(c) 
establishes its dynamical stability.


\onecolumn

\twocolumn


\begin{figure}[!t]

\begin{center}

\includegraphics[trim = 0mm 0mm 0cm 0mm, clip,width=.9\linewidth,clip]{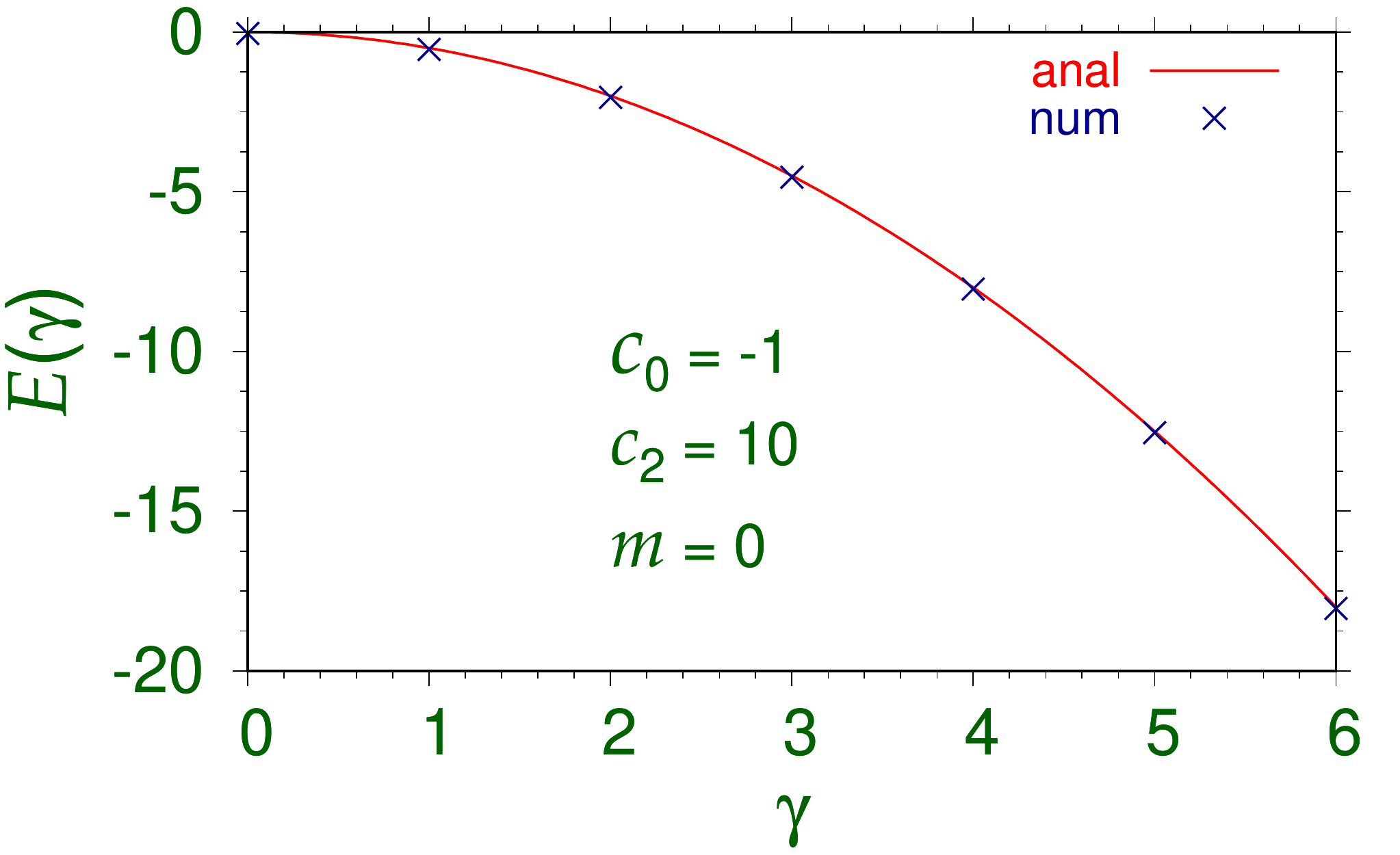}

\caption{(Color online) Numerical (num)     result  of energy $E(\gamma)$  of the stationary   vector soliton with $c_0=-1, c_2=10, m=0$ as a function of $\gamma$ compared with  the analytic (anal) result 
  $[E(\gamma)=-\gamma^2/2-c_0/24$]
obtained with the secant hyperbolic function, viz. Eq. (\ref{En2}).  }
\label{fig6} \end{center}
\end{figure}

Next we test the reliability of the analytic expression for energy of the vector soliton (\ref{En2})  by comparing 
it with the energies obtained numerically by the imaginary-time propagation method. The result of this study is 
exposed in Fig. \ref{fig6}, where we display the  energy $E(\gamma)$ for different  $\gamma$  as obtained 
numerically  and from the analytic approximation   (\ref{En2}).  The good agreement between the two results is 
assuring.

\begin{figure}[!t]
\begin{center}
\includegraphics[trim = 0mm 0mm 0cm 0mm, clip,width=.9\linewidth,clip]{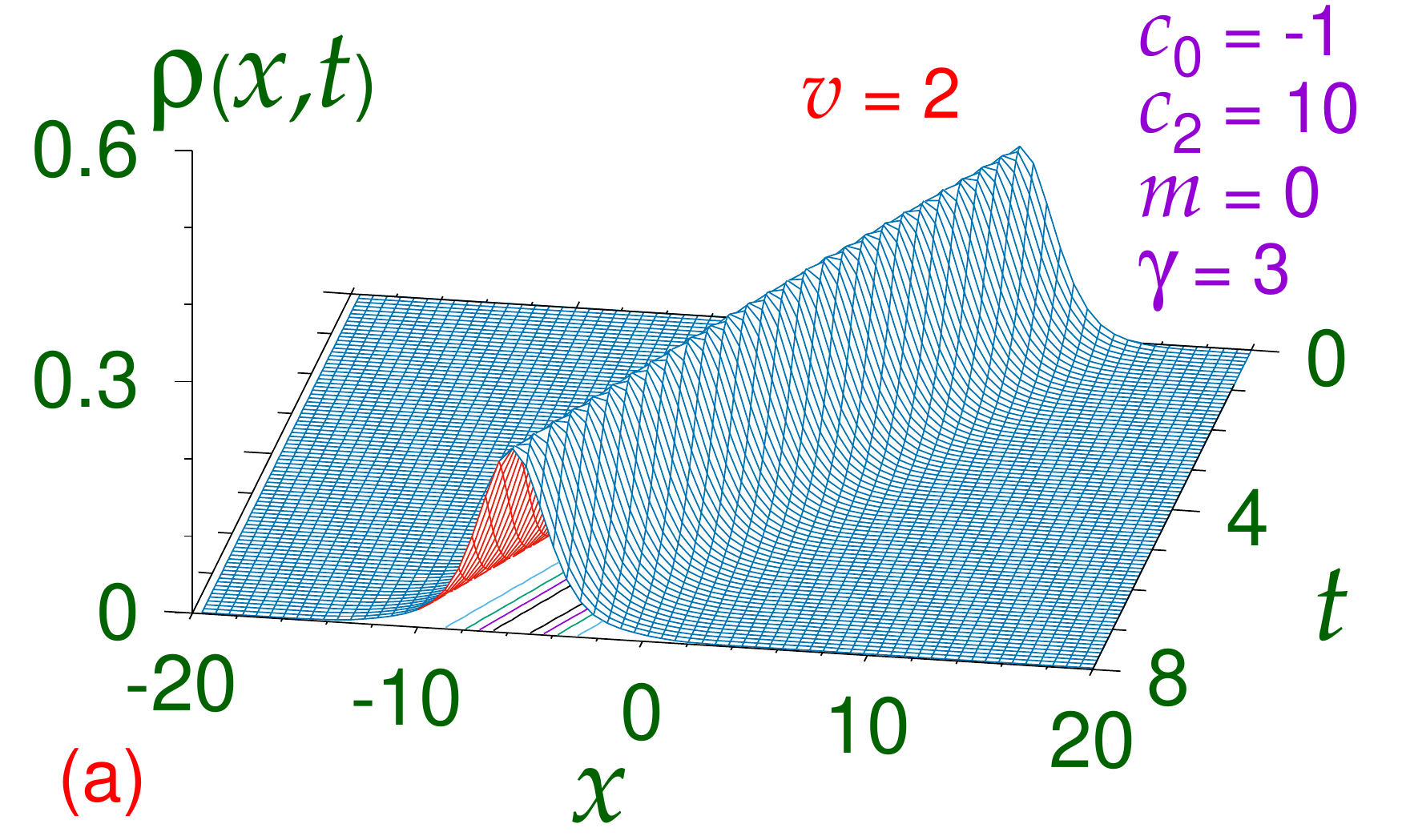}
\includegraphics[trim = 0mm 0mm 0cm 0mm, clip,width=.9\linewidth,clip]{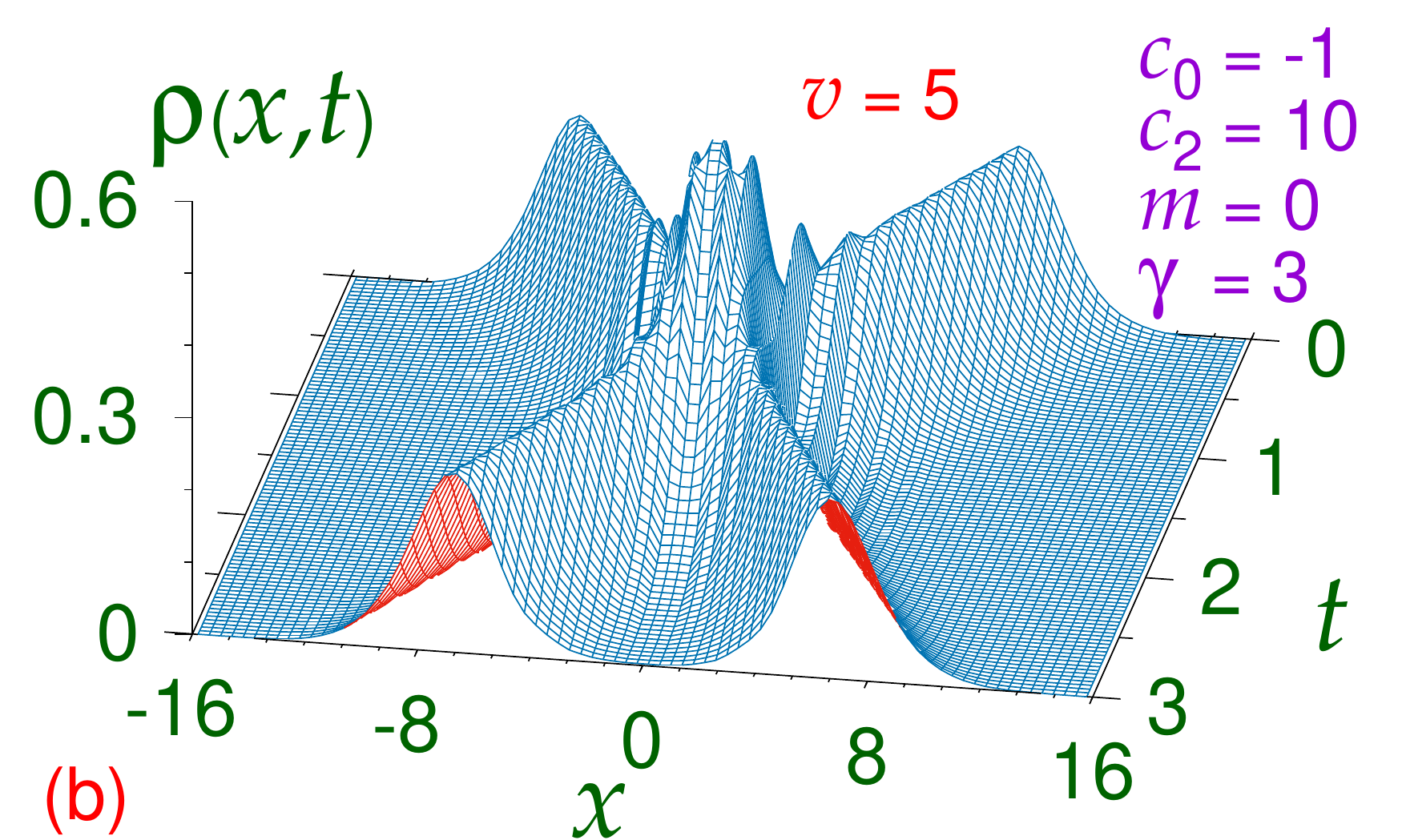}
\includegraphics[trim = 0mm 0mm 0cm 0mm, clip,width=.9\linewidth,clip]{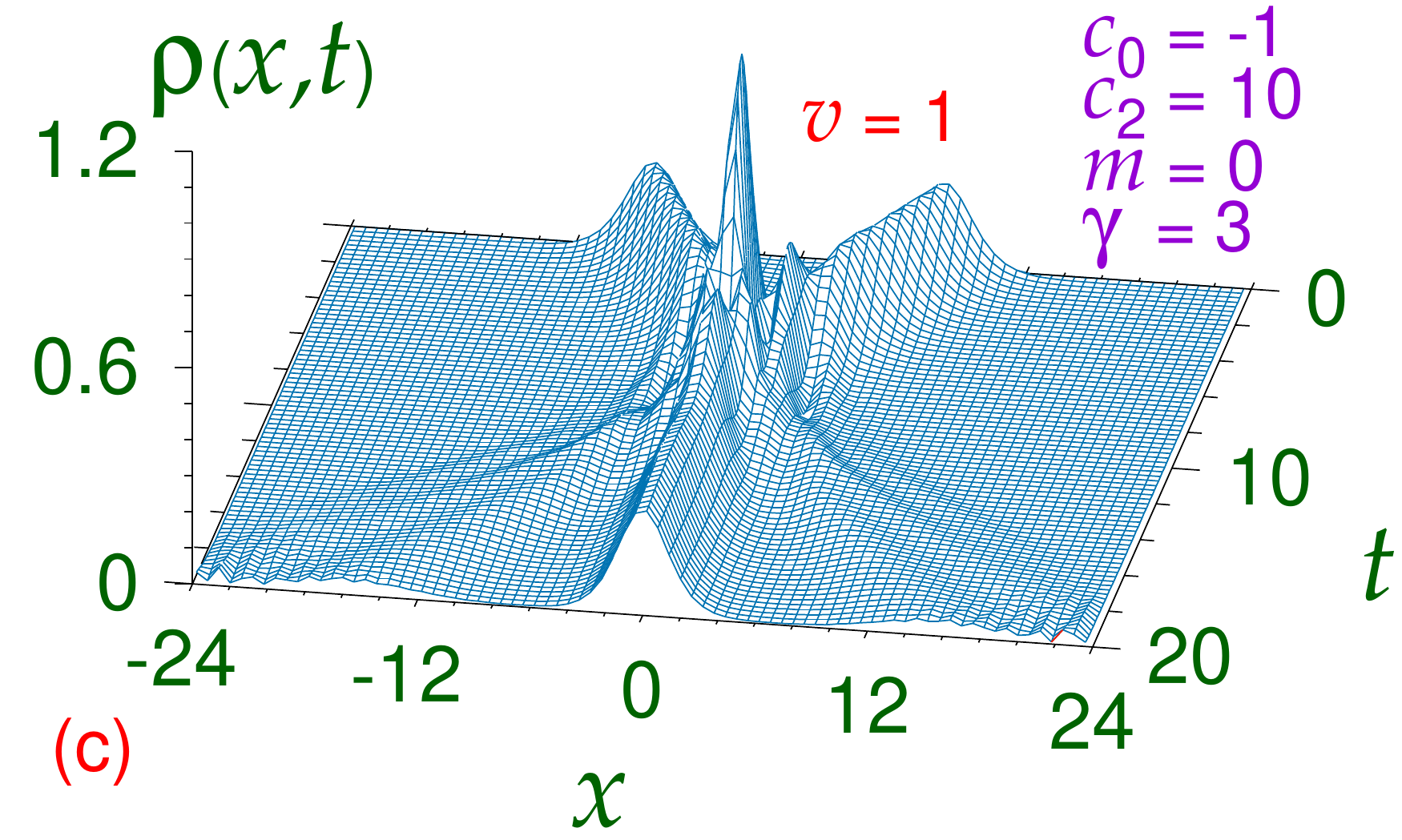}

\caption{(Color online) (a) Steady propagation by real-time simulation with velocity $v=2$ of a vector soliton preserving its shape. The initial imaginary-time wave function placed at $x=10$ is  multiplied  by $\exp ivx)$ to start the motion.
(b) Elastic nature of collision  dynamics of two  vector  solitons of (a)    
illustrated through a plot of total density $\rho(x,t)$, obtained by real-time propagation, 
versus $x$ and $t$. At $t=0$ individual solitons were placed at $x=\pm 8$ and set into motion 
in opposite directions with velocity $v=\pm 5$, respectively. (c)  Inelastic nature of the  same dynamics with $v=\pm 1$. Other parameters were $c_0=-1, c_2=10, m=0, \gamma =3$.}
\label{fig7} \end{center}
\end{figure}

The dynamics of the moving solitons  is next studied by  first generating
a  vector  soliton numerically using imaginary-time propagation. 
The complex wave-function components so obtained are then multiplied by a complex phase $\exp(ixv)$,
which is used as the initial state in real time simulation to obtain a moving soliton with velocity $v$ 
in the limit of vanishing small space and time steps $dx$ and $dt$. It is known that the SO-coupled 
GP equations (\ref{gp1})  and (\ref{gp3}) are not Galilean invariant \cite{sol1d}. The wave-function ansatz 
(\ref{aussian}) or (\ref{aussian2}) is a linear combination of two fundamental degenerate solutions of the SO-coupled linear  GP 
equations  (\ref{GP1}) and (\ref{GP3}). In the rest frame $(v=0)$
the physical solution (\ref{aussian}) or (\ref{aussian2}) is constructed as a linear combination of these two fundamental 
solutions.  As the underlying GP equation is not Galilean invariant, for a moving soliton ($v\ne 0$)
the two fundamental degenerate solutions will have different energies  \cite{sol1d} and one can not consider the same 
mixture of these two fundamental  solutions  to construct the same physical solution. Hence the vector 
soliton cannot move maintaining the same density profile of the component solitons. This will lead to a
spin-mixing dynamics during motion with constant change of density profile of the components
by a periodic transfer of atoms between components \cite{sol1d}, however, preserving the total density profile $\rho(x)$, at 
least for a reasonable interval of time. 
First we demonstrate that the vector soliton can move maintaining the total density profile, although the component 
densities are not conserved.   
  In Fig. \ref{fig7}(a) we show the total density $\rho(x,t)$, obtained by real-time simulation,
of the vector soliton  with parameters  $c_0=-1, c_2=10, m=0$ and $\gamma=3$   moving with a velocity  $v=2$, initially 
placed at  $x=10$.  A steady propagation of the vector soliton in this figure is convincing. 

 To demonstrate the solitonic property of the   vector soliton, we 
study the collision of two vector solitons each generated by imaginary-time propagation with parameters
$c_0=-1, c_2=10, m=0, \gamma=3$.   We take two vector solitons and place them at positions $x\equiv d =\pm 8$ 
and set them in motion in opposite directions    with velocity $v=\pm 5$  so as to collide  at 
$x=0$ after time $t=d/v=1.6$. The solitons are found to interact and pass through each other.  
The total density of each soliton is conserved after collision, showing
its elastic nature.  This is displayed in Fig. \ref{fig7}(b) via a plot of total density $\rho(x,t) $ 
 of the two solitons during collision. The two solitons emerge with the same velocity  and the same total 
density after collision.  However, at a small velocity, the collision of the two solitons turns inelastic.  
This is exhibited in Fig. \ref{fig7}(c), where we plot the total density of two colliding solitons  initially 
placed at $x=\pm 8$ 
 moving  in opposite directions
with a velocity $v=\pm 1$.   In this case most atoms of the two vector solitons combine to form a larger one which  stay at rest at $x=0$  while smaller fractions of atoms move in  approximate forward directions.  The collision is inelastic with a loss of identity of the two vector solitons. 
Similar panorama takes place at  smaller incident velocities.

\section{Summary}
\label{Sec-IV}

We studied the generation,   and collision dynamics   of three-component  $(F_z=0,\pm 1)$ 1D 
 vector solitons of a
SO-coupled  spin-1
polar BEC ($c_2> 0$) by a numerical solution and 
an analytic approximation of the mean-field GP equation.  The SO coupling is taken as  $\gamma p_x \Sigma_y$, which is an equal-strength  mixture of Rashba and Dresselhaus SO couplings \cite{Lin}.   The solitons appear for  interaction strength   $c_0<0$.  
The vector solitons could have a very large number of maxima and minima. 
The vector solitons with more than  150 maxima and minima 
were demonstrated to be stable by real-time propagation during a large interval of time.  
  At large velocities, 
the collision dynamics between two such vector solitons is found to be elastic with the conservation of the 
total densities of each individual vector solitons, viz. Fig. \ref{fig7}(b). However, the individual densities of the components are not conserved 
during collision.   The collision with a small velocity  is inelastic with the destruction 
of the individual solitons, viz. Fig. \ref{fig7}(c). A remnant vector soliton soliton molecule appears at rest 
with lot of dissipation of atoms.  With present know-how an experiment can be performed 
to test the predictions of the present theoretical study.


\section*{Acknowledgements}
This work is financed by the Funda\c c\~ao de Amparo \`a Pesquisa do Estado de 
S\~ao Paulo (Brazil) under Contract Nos. 2013/07213-0, 2012/00451-0 and also by 
the Conselho Nacional de Desenvolvimento Cient\'ifico e Tecnol\'ogico (Brazil).


\end{document}